\documentclass[prepint,12pt]{elsarticle}
\usepackage[utf8]{inputenc}

\usepackage{amssymb}
\usepackage{amsmath}
 \usepackage{amsthm}
\usepackage{braket}

\makeatletter
\def\ps@pprintTitle{%
\let\@oddhead\@empty
\let\@evenhead\@empty
\def\@oddfoot{}
\def\@evenfoot{}}
\makeatother

\begin{document}

\begin{frontmatter}


\title{New multiple soft theorem from split-helicity amplitude decomposition}


\author[a]{Andriniaina N. Rasoanaivo}\ead{andriniaina@aims.ac.za}
\author[b]{Fanomezantsoa A. Andriantsarafara}
\address[a]{Sciences Expérimentales et des Mathématiques, Ecole Normale Supérieure, \\ Université d'Antananarivo - Madagascar.\vspace*{.3cm}}
\address[b]{Mention Physique et Applications, Faculté des Sciences, \\ Université d'Antananarivo - Madagascar.\vspace*{.3cm}}

\begin{abstract}
We investigate the soft decomposition of tree-level gluon amplitudes with split-helicity configurations. First, we show how any split-helicity amplitude can be fully fixed from inverse soft limit using BCFW calculation. We show how the double and triple soft theorems can manifest beyond perturbative calculations through split-helicity decomposition. Next, we obtain a compact formula for the corresponding soft operator which is a simple combination of individual single soft operators. Then we generate the leading and sub-leading soft factors using soft momentum expansion, to compare with results derived from standard calculation.

\end{abstract}




\end{frontmatter}



\section{Introduction}

Weinberg's soft theorem describes how scattering amplitudes behave when the energy of a massless particle involved in the process tends to zero. In this limit, the low energy dependence of amplitudes tends to factorize out from the high energy contributions. The leading order of the soft dependence is known as the Weinberg soft factor \cite{Weinberg:1965nx}, and its full contribution is shown to be an operator that acts on the high energy regime of a lower point amplitude.

One of the main features of the soft factorization is its universality where the expression of the soft operator depends mainly on the nature of the soft particle \cite{Higuchi:2018vyu}. This universality suggests that the soft operator should be governed by a set of laws and it can be derived directly from physical properties of the soft particle. In \cite{Campoleoni:2017mbt}, the leading soft factors for spin 1 and 2 are derived directly from the four-dimensional higher spin asymptotic symmetries. In \cite{Rasoanaivo:2020yii, Rabearinoro:2022jwi}, the leading and sub-leading soft factors are both derived by solving a set of equations from helicity constraints which is a direct implication of Lorentz transformation.

The universality of the theorem also tends to consider the soft operator as an independent physical quantity with which the idea of an inverse soft limit is studied to restore soft particles in the hard process \cite{Boucher-Veronneau:2011rwd, Rodina:2018pcb}. 
One of the main objectives of the inverse soft limit approach is to build higher point amplitudes from lower points by adding particles one by one. However, adding a particle softly through ISL does not always lead to an amplitude. To counter such problem, it is shown in \cite{Rodina:2018pcb} that multiple ISL is needed in order to reconstruct higher point amplitude. Such reconstruction can be viewed as a decomposition of amplitudes to the the possible soft factorization of single soft particle \cite{Rasoanaivo:2022mgv}. 

In this work, we would like to study of multiple soft factorization behavior from soft decomposition. The study of multiple soft shows that the leading and sub-leading soft factor might be expressed in terms of the individual soft factors. For the case of double soft of adjacent particle, the factorization depends main;y on the order at which the soft limit was taken. However if the two particle go soft simultaneously it is hard to express the soft contribution with the two single soft factorizations \cite{Klose:2015xoa,Volovich:2015yoa, Georgiou:2015jfa}. In the quest to understand multiple soft operators, one can ask the following question: \emph{would a soft decomposition be able to make multiple soft factorization manifest?}

To answer this question, we are investigate the soft decomposition of split-helicity amplitudes of gluons. In this proof, we will discuss the equivalence between the BCFW shifts and soft shifts as developed in \cite{Boucher-Veronneau:2011rwd}. We then show that with the soft decomposition, we can explicitly derive split-helicity amplitude from multiple ISL, and such decomposition also simplify factorization of multiple soft gluon. This approach will provide a well-defined non-pertubative soft operator to generate the double and triple soft factor at any order respectively for two and three adjacent gluons with non identical helicity configuration.  

\section{Single soft operators }

In this section, we are giving a short review of the single soft limit of gluon amplitudes. We consider a $(n+1)$ point amplitude in which the energy of the $(n+1)$-th gluon is taken to be soft. In this limit, the Weinberg's theorem states that amplitude factorizes into an operator acting on a lower-point amplitude as follows
\begin{equation}
A_{n+1}(p_1,p_2,\ldots,p_n,\epsilon p_{n+1})\xrightarrow{\quad\epsilon\to0\quad}\hat{S}(\epsilon p_{n+1}) A_n(p_1,p_2,\ldots,p_n),
\label{soft_theorem}
\end{equation}
here $\epsilon$ is a parameter to tune the energy of the soft particle, and  $\hat{S}(p_{n+1})$ is the soft operator associated to the $p_{n+1}$-th gluon. Such factorization is well understood and shown to be independent of the nature of the process but dependent mainly on the nature of the soft particle. One can use the action of the little group \cite{Britto:2004ap, Elvang:2013cua}, which is a subgroup of Lorentz that leaves momentum invariant, on amplitudes to derive the equations that describe the soft operator as introduced in \cite{Rasoanaivo:2020yii}. In terms of the algebra of the little group, helicity operator $\hat{H}(p_j)$, the helicity constraints of the soft operator are given by the following commutation relations
\begin{equation}
\left [\hat{S}(p_i),\hat{H}(p_j)\right ]=h_i\delta_{ij}\hat{S}(p_i).
\label{helicityconstraints}
\end{equation}
To simplify this equation, it is better to use spinor helicity variables, where the momentum of massless particle is represented by two spinors: one left-handed $\lambda_a$ and the other right-handed $\bar{\lambda}_{\dot{a}}$. Such change of variables is a direct consequence of the isomorphism between the Minkowskian momentum space and the spinor space where for massless particle the incidence relation is given by $p_\mu \sigma_{a\dot{a}}^\mu=\lambda_a\bar{\lambda}_{\dot{a}}$ as in \cite{Britto:2004ap,Elvang:2013cua}. 
In this representation, the helicity constraints \eqref{helicityconstraints} of the soft factor, which is the leading soft contribution, is reduced into linear partial differential equations under two boundary conditions. The first condition comes from the fact that soft factorization only occurs at low energy so the soft factor has to go to zero at high energy limit. The second condition is include only single pole divergences at a low energy limit which comes from the locality of the theory. Combined with the energy-momentum conservation, the single soft operator of a particle labeled by $k$, adjacent to two particles labeled respectively by $i$ and $j$, is given by

\begin{equation}
\hat{S}_{i,j}(k^+)=\frac{\braket{ij}}{\braket{ik}\braket{kj}}\hat{U}_k^+(i,j) \quad\text{and}\quad 
\hat{S}_{i,j}(k^-)=\frac{[ij]}{[ik][kj]}\hat{U}_k^-(i,j),
\label{soft_operator}
\end{equation}
where the $+$ and $-$ design the heliciy of the gluon, here $\hat{U}_k^{\pm}(i,j)$ are soft momentum shift operators acting on the $i$-th and $j$-th particles. With the spinor variables, the momentum shifts are given by

\begin{equation}
\hat{U}_k^+(i,j):
\left \{\begin{aligned}
&\bar{\lambda}_i^{\dot{a}}\to \bar{\lambda}_i^{\dot{a}}+\frac{\braket{jk}}{\braket{ji}}\bar{\lambda}_k^{\dot{a}}\\
&\bar{\lambda}_j^{\dot{a}}\to \bar{\lambda}_j^{\dot{a}}+\frac{\braket{ik}}{\braket{ij}}\bar{\lambda}_k^{\dot{a}}
\end{aligned}\right .
\quad\text{and}\quad \hat{U}_k^-(i,j):
\left \{\begin{aligned}
&{\lambda}_i^{a}\to {\lambda}_i^{a}+\frac{[jk]}{[ji]}{\lambda}_k^{a}\\
&{\lambda}_j^{a}\to {\lambda}_j^{a}+\frac{[ik]}{[ij]}{\lambda}_k^{{a}}
\end{aligned}\right .
\label{soft-shifts}
\end{equation}
In the soft limit it is important to mention that the soft factor contribution and the momentum shifts occur simultaneously that leads to be commutation of $S^{(0)}$ and $\hat{U}$, and as presented in \cite{Boucher-Veronneau:2011rwd, Klose:2015xoa}, the soft operator can be expended as
\begin{equation}
\hat{S}_{i,j}(k)={S}^{(0)}_{i,j}(k)+{S}^{(1)}_{i,j}(k)+{S}^{(2)}_{i,j}(k)+\cdots
\end{equation}

\section{Split-helicity amplitude}

In the spinor helicity approach, maximally violating helicity amplitudes are known to have a simple compact expression, which is described by Park-Taylor formula \cite{Parke:1986gb}. From the inverse soft perspective, we will show that the split-helicity amplitudes have the simplest non-trivial ISL recursion, which can be derived directly from BCFW recursions. Such an algorithm unravel the soft structure with which we can decompose the amplitude into the possible way of soft factorization. 

\subsection{Soft decomposition}
\label{def:dipole}

A split-helicity configuration amplitudes are ordered amplitudes in which the ordering of particles are arranged in such a way the negative (and/or positive) helicity are grouped in one region \cite{Britto:2005dg}. Since the ordered amplitude are cyclic invariant \cite{Elvang:2013cua}, we can always labelled split-helicity amplitudes of $m$ negative helicity particles as 
\begin{equation}
A_{n,m}^\text{split}=A_n(1^-,\ldots,m^-,(m+1)^+,\ldots,n^+).
\label{n-point}
\end{equation}
In order to compute such an amplitude using the BCFW recursion, we will use the $[1,n\rangle$ BCFW shifts in which the right and left handed spinors respectively to the particle labelled by $n$ and $1$ are shifted in such a way they remain on-shell and the total momentum is conserved 
\begin{equation}
\left \{\begin{aligned}
\bar{\lambda}_1&\to \hat{\bar{\lambda}}_1=\bar{\lambda}_1-z\bar{\lambda}_n\\
\lambda_n&\to\hat{\lambda}_n= \lambda_n+z\lambda_1
\end{aligned}\right .
\label{BCFW-shifts}
\end{equation}
Since any $n$ gluon amplitude, for $n\geq4$, with only one positive or only one negative helicity vanishes therefore the only two BCFW terms that contribute to the computation of $A_{n,m}^\text{split}$ are given by 
\begin{equation}
\begin{aligned}
A_{n,m}^\text{split}&=\underbrace{\begin{aligned}\text{\includegraphics[scale=.3]{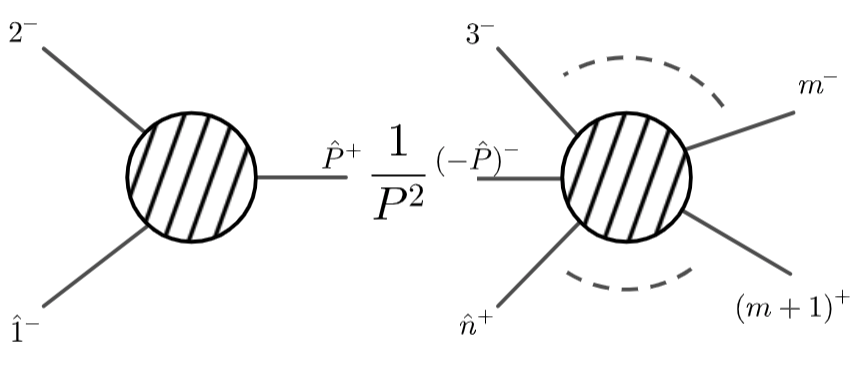}}\end{aligned}}_{\large\mathcal{X}}+\underbrace{\begin{aligned}\text{\includegraphics[scale=.3]{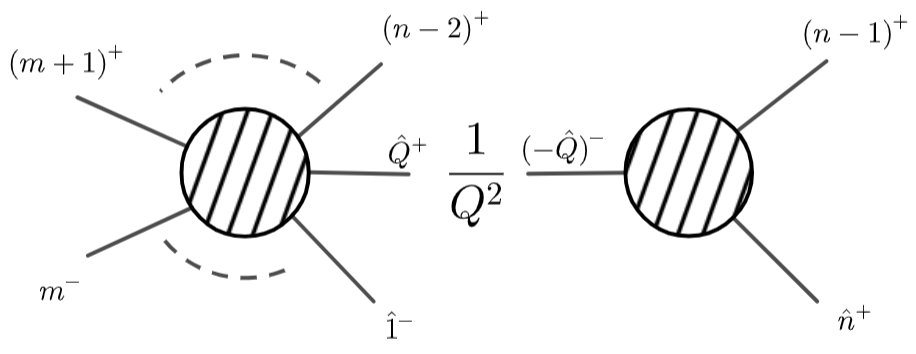}}\end{aligned}}_{\large\mathcal{Y}}
\end{aligned}
\label{dipole_nm}
\end{equation}

In order to simplify these BCFW terms, it is useful to consider separately the two BCFW contribution to connect them with their respective soft factorization.  For the first contribution, we will first show that the three point left amplitude can be reduced into soft factor 
\begin{equation}
\mathcal{X}=\frac{A_3(\hat{1}^-,2^-,\hat{P}^+) }{P^2}\times A_{n-1}(-\hat{P}^-,3^-,\ldots, m^-,(m+1)^+,\ldots, \hat{n}^+)
\label{contribution1}
\end{equation}
By combining the momentum conservation, applied on the left and right sub-amplitude of the first contribution, and the onshell condition of the recursion allow us to constrain the momentum $\hat{P}$ and fix the variable $z$ of the BCFW shift as $z=[12]/[n2]$. Hence, the shifted momenta $\hat{P}$ and $\hat{p}_n$ are then given the following expression

\begin{equation}
-\hat{P}
=\left (\lambda_2+\frac{[{n1}]}{[{n2}]}{\lambda}_1\right )\bar{\lambda}_2
\quad\text{and}\quad \hat{p}_n=\left (\lambda_n+\frac{[{21}]}{[{2n}]}{\lambda}_1\right )\bar{\lambda}_n.
\label{shift-P}
\end{equation}
Using the relation above, the left sub-amplitude of $\mathcal{X}$ can be reduced to a single soft factor 
\begin{equation}
\frac{A_3(\hat{1}^-,2^-,\hat{P}^+)}{P^2}=\frac{\braket{12}^4}{\braket{2\hat{P}}\braket{\hat{P}1}\braket{12}}\times\frac{1}{\braket{12}[21]}=\frac{[{n2}]}{[{n1}][{12}]}
\label{A3X}
\end{equation}
While the BCFW shift $\hat{P}$ and $\hat{p}_n$ in the right sub-amplitude of $\mathcal{X}$ can be identify with a soft shift $U_1^-(n,2)$  of $p_2$ and $p_n$ acting only on ${\lambda}_2$ and ${\lambda}_n$ as define in \eqref{soft-shifts}, 
\begin{equation}
-\hat{P}=U_1^-(n,2)\lambda_2\bar{\lambda}_2 \quad\text{and}\quad \hat{p}_n=U_1^-(n,2)\lambda_n\bar{\lambda}_n,
\end{equation}
which leads to the following relation

\begin{equation}
\begin{aligned}
A_{n-1}(-\hat{P}^-,
\ldots, m^-,&(m+1)^+,\ldots, \hat{n}^+)=\\
&U_1^-(n,2)A_{n-1}(2^-,\ldots, m^-,(m+1)^+,\ldots, n^+).
\end{aligned}
\label{shifts_equal}
\end{equation}
Combining \eqref{A3X} and \eqref{shifts_equal} together, the $\mathcal{X}$-BCFW term in \eqref{contribution1} can be written as a soft factorization of the $n$-point amplitude \eqref{n-point} in which the momentum $p_1$ in considered to be soft,

\begin{equation}
\mathcal{X}=\hat{S}_{n,2}(1^-)A_{n-1}(2^-,\ldots, m^-,(m+1)^+,\ldots, n^+).
\label{soft1}
\end{equation}
Similarly, we can apply the same approaches as for $\mathcal{X}$ to the second contribution of the BCFW term in \eqref{dipole_nm},
\begin{equation}
\mathcal{Y}=  A_{n-1}(\hat{1}^-,\ldots, m^-,(m+1)^+,\ldots,(n-2)^+,\hat{Q}^+)\times\frac{A_3(-\hat{Q}^-,({n-1})^+,\hat{n}^+)}{Q^2},
\label{contribution2}
\end{equation}
where the BCFW shifts leads to

\begin{equation}
\hat{Q}={\lambda}_{n-1}\left (\bar{\lambda}_{n-1}+\frac{\braket{1,n}}{\braket{1,n-1}}\bar{\lambda}_n\right )
\quad\text{and}\quad \hat{p}_1={\lambda}_1\left (\bar{\lambda}_1+\frac{\braket{n-1,n}}{\braket{n-1,1}}\bar{\lambda}_n\right ).
\label{shift-Q}
\end{equation}
With that in hand, the three-point amplitude in \eqref{contribution2} will be reduced to a soft factor, while the BCFW shifts \eqref{shift-Q} will be connected with a soft deformation as in \eqref{soft-shifts}, 
  
\begin{equation}
 \left\{\begin{aligned}
 &\frac{A_3(-\hat{Q}^-,({n-1})^+,\hat{n}^+)}{Q^2}= \frac{\braket{n-1,1}}{\braket{n-1,n}\braket{n,1}},\\
 &\hat{Q}=U_n^+(n-1,1){\lambda}_{n-1}\bar{\lambda}_{n-1},\\[.2cm]
 &\hat{p}_1=U_n^+(n-1,1)\lambda_1\bar{\lambda}_1.\\[.1cm]
 \end{aligned}\right .
 \end{equation} 
By taking the last relations into \eqref{contribution2}, allows us to express the $\mathcal{Y}$-BCFW as a soft factorization of the $n$-point amplitude \eqref{n-point} in which the momentum $p_n$ in considered to be soft, 

\begin{equation}
\mathcal{Y}=\hat{S}_{n-1,1}(n^+)A_{n-1}({1}^-,\ldots, m^-,(m+1)^+,\ldots, (n-1)^+).
\label{soft2}
\end{equation}
Looking at the soft factorization expressions equivalence of each BCFW terms $\mathcal{X}$ and $\mathcal{Y}$ of the split-helicity amplitude \eqref{dipole_nm}, it is important to notice that the $A_{n-1}$ sub-amplitudes are also split-helicity amplitudes. For that an $A_{n-1,m-1}^\text{split}$ split amplitude contributes to the sub-amplitude of $\mathcal{X}$ at \eqref{soft1} and $A_{n-1,m}^\text{split}$ to the sub-amplitude of $\mathcal{Y}$ at \eqref{soft2}. Combine the two soft factorization expressions back to the BCFW expression we obtain the following inverse soft recursion relation

\begin{equation}
\begin{aligned}
A_{n,m}^\text{split}=\hat{S}_{n,2}(1^-)A_{n-1,m-1}^\text{split}+\hat{S}_{n-1,1}(n^+)A_{n-1,m}^\text{split}.
\end{aligned}
\label{inverse_soft_recursion}
\end{equation}

\subsection{Interpretation}

This result shows how the BCFW recursion for any split-helicity configuration of the type \eqref{n-point} can be reconstructed recursively by softly adding gluon respectively to two lower-point split configurations. Here a positive helicity gluon is added to a $(n-1)$-split of $m-1$ positive helicities while a negative helicity is added to a $(n-1)$-split of $m$ positive helicities, then combine to form the $n$-point split-helicity configuration of $m$ positive helicities. The relation \eqref{inverse_soft_recursion} also show the decomposition of a $n$-split configuration to the different possible outcomes, either $(n-1)$-split of $m$ or $m-1$ positive helicities, in the soft theorem. As mention in \cite{Rasoanaivo:2022mgv}, such soft decomposition is not unique but it reflects the BCFW shifts from which the soft decomposition was derived. Here we choose to shift \eqref{BCFW-shifts} for $p_1$ and $p_n$ that is why the amplitude is decomposed with the soft channels associated to $p_1$ and $p_n$. 

The soft recursive relation of $n$-split helicity configuration is similar to the case of maximal helicity violating (MHV) amplitude, helicity configuration with only two negative helicity gluons, where the amplitude can be reconstructed by softly adding one positive helicity gluon to a lower point MHV amplitude. Here the relation shows how the amplitude can be softly reconstructed from multiple soft factorization as proposed in \cite{Rodina:2018pcb}. Such relation can be used to reconstruct the amplitude from lower-point amplitude, diagrammatically represented by the following relation
\begin{equation}
\begin{aligned}
\text{\includegraphics[scale=.55]{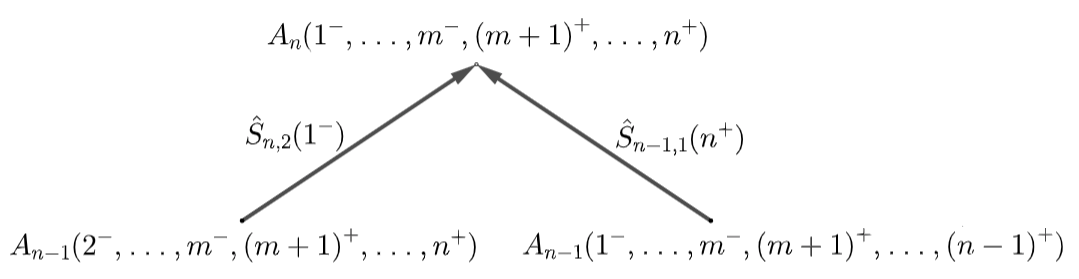}}
\end{aligned}
\label{inverse_soft_recursion_tree}
\end{equation}

\section{Multiple soft operator}

In this section, we will focus on understanding the soft factorization of amplitudes when more than one gluon goes soft beyond the perturbative derivation. Such factorization depends mainly on the order in which gluons become soft and also depends on the helicity configuration of the gluons. 
In fact for the case where the gluons are getting soft consecutively, i.e. go soft at different time, the corresponding soft operator will be the product of the respective single soft operators taken in that consecutive order. A more interesting situation is when the gluons are getting soft simultaneously. However, as discussed in \cite{Klose:2015xoa, Volovich:2015yoa}, the results for simultaneous soft gluons with the same helicities are the same as those for consecutive soft calculations. This is why we are focusing our study on the simultaneous soft gluons with mixed helicity configuration.

At first we will derive and study the double soft operator in the case where the helicity of the two gluons are different ($1^+2^-$) using the soft decomposition \eqref{inverse_soft_recursion} on a six point split-helicity amplitude. Then second we will use similar idea using a seven point split-helicity amplitude to derive triple soft operator for the case of helicity configuration $(1^+2^-3^-)$.
 
\subsection{Double soft}
\begin{figure}
\centering
\includegraphics[scale=.5]{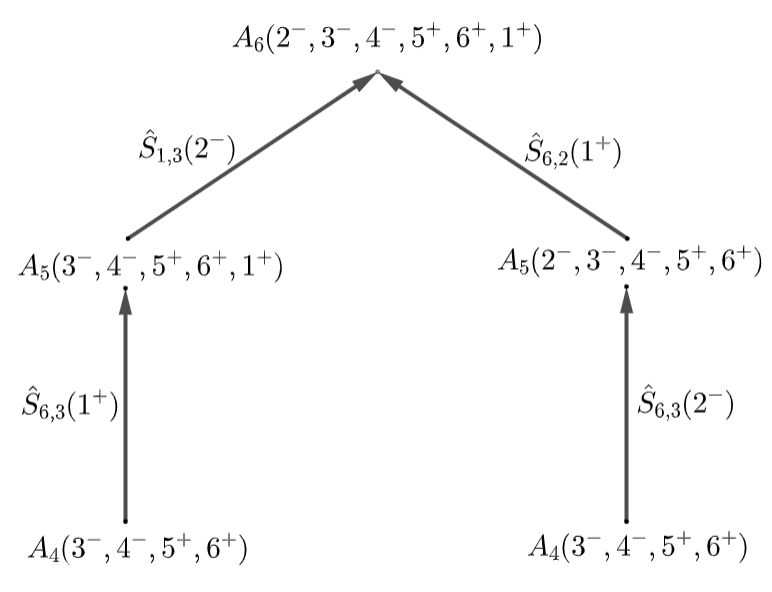}
\caption{Soft reconstruction of a six point split-helicity amplitude.}
\label{diagram_A6}
\end{figure}

Since split-helicity amplitudes do not depend on the way they are labeled but on how the negative helicities are grouped in its configuration. 
Therefore, to derive the non-perturbative double soft operator, we consider the following six point split-helicity amplitude 
\begin{equation}
 A_{6,3}^\text{split}=
A_6(2^-,3^-,4^-,5^+,6^+,1^+).
 \end{equation} 
Starting with the inverse soft recursion relation \eqref{inverse_soft_recursion_tree}, we can decompose the six point split-helicity amplitude above into five point split-helicity amplitudes which are MHV and $\overline{\text{MHV}}$ amplitudes. Then using the fact that MHV (resp $\overline{\text{MHV}}$) amplitude can be reconstructed by softly adding one positive (resp negative) helicity gluon to a lower point amplitude. Such reconstruction of the six point amplitude can be diagrammatically represented as in \eqref{inverse_soft_recursion_tree} with the diagram in fig.\ref{diagram_A6}, which can be analytically written as
\begin{equation}
\begin{aligned}
A_6(2^-,3^-,4^-,&5^+,6^+,1^+)=\\&\Big [\hat{S}_{6,2}(1^+)\hat{S}_{6,3}(2^-)+\hat{S}_{1,3}(2^-)\hat{S}_{6,3}(1^+)\Big ]A_4(3^-,4^-,5^+,6^+).
\end{aligned}
\label{non_perturbative_factorization}
\end{equation}
The relation above \eqref{non_perturbative_factorization} can be viewed as a soft factorization, even the expression is a decomposition of the six point into different soft channel at high energy, i.e. non of the momentum is taken to be soft in the expression. In the relation, the six point split-helicity amplitude $A_{6,3}^\text{split}$ into a term composed with single soft operators associated with momentum $p_1$ and momentum $p_2$ which is factorized from a four point amplitude. In the limit these momenta 1 and 2 are simultaneously soft, the term composed with them single soft operators is exactly the double soft associated defined by
\begin{equation}
\hat{S}_{6,3}(1^+,2^-)=\hat{S}_{6,2}(1^+)\hat{S}_{6,3}(2^-)+\hat{S}_{1,3}(2^-)\hat{S}_{6,3}(1^+).
\label{double_operator}
\end{equation}

\subsubsection{Perturbative comparison}
First, we will use the expression of the single soft operator defined in \eqref{soft_operator} to obtain the explicit expression of the non-perturbative double soft operator \eqref{double_operator}. Then, we will apply moment expansion as the particles $1$ and $2$ are getting soft.
In that regard the first term of the operator \eqref{double_operator} can be written as 
\begin{equation}
\begin{aligned}
\hat{S}_{6,2}(1^+)\hat{S}_{6,3}(2^-)&=\frac{\braket{62}}{\braket{61}\braket{12}}\hat{U}^+_{6,2}(1)\frac{[63]}{[62][23]}\hat{U}^-_{6,3}(2)\\
&=\frac{\braket{62}^2}{\braket{61}\braket{12}}\frac{[3(1+6)2\rangle}{\langle 6(1+2)3]}\frac{1}{s_{126}}\hat{U}^+_{6,2}(1)\hat{U}^-_{6,3}(2),
\end{aligned}
\end{equation}
where $s_{126}=(p_1+p_2+p_6)^2$ is the generalized Mandelstam variable with three momenta, and $\langle i(j)k]=\braket{ij}[jk]$ is a spinor contraction as introduced in \cite{Britto:2004ap}. By taking $1$ and $2$ to be soft, we expand each term in that expression up to the next leading soft. To keep track the expansion order we will use the same counting as in \cite{Volovich:2015yoa} where we assign $\sqrt{\epsilon}$ for each $\lambda$ and $\bar{\lambda}$ of the soft momenta. 

In the expression, we have

\begin{equation}
\frac{\braket{62}^2}{\braket{61}\braket{12}}\frac{[3(1+6)2\rangle}{\langle 6(1+2)3]}=
\underbrace{\frac{\braket{62}^3[36]}{\braket{61}\braket{12}}\frac{1}{\langle 6(1+2)3]}}_{1/\epsilon}+
\underbrace{\frac{\braket{62}^2[31]}{\braket{61}}\frac{1}{\langle 6(1+2)3]}}_{\epsilon^0}.
\end{equation}
Then next, the Mandelstam $s_{126}$ in which we use $q_{12}=p_1+p_2$ and also use the on-shell condition $p_6^2=0$. Here the momentum $q_{12}$ is in the order of $\epsilon$ and the soft expansion of the term leads to following up to the order of $\epsilon^0$
\begin{equation}
\begin{aligned}
\frac{1}{s_{126}}&=\frac{1}{(p_1+p_2+p_6)^2}
&=\frac{1}{2 p_6 \cdot q_{12}}\left(1-\frac{\braket{12}[21]}{2 p_6 \cdot q_{12}}
\right)+\mathcal{O}(\epsilon).
\end{aligned}
\end{equation}
Since the momentum shifts, defined in \eqref{soft-shifts}, are soft deformation of several momenta, then them action can be expanded up to the next leading order which is of order $\epsilon$ such that
\begin{equation}
\left \{\begin{aligned}
&\hat{U}^+_{6,2}(1)= 1+\frac{\braket{21}}{\braket{26}} \bar{\lambda}^{\dot{a}}_1\frac{\partial}{\partial \bar{\lambda}^{\dot{a}}_6} +\frac{\braket{61}}{\braket{62}} \bar{\lambda}^{\dot{a}}_1\frac{\partial}{\partial \bar{\lambda}^{\dot{a}}_2}+\mathcal{O}(\epsilon^2)\\
&\hat{U}^-_{6,3}(2)=1+\frac{[62]}{[63]} {\lambda}^{{a}}_2\frac{\partial}{\partial {\lambda}^{{a}}_3} +\frac{[32]}{[36]} {\lambda}^{{a}}_2\frac{\partial}{\partial {\lambda}^{{a}}_6} +\mathcal{O}(\epsilon^2)
\end{aligned}\right .
\end{equation}
By keeping of the order of soft expansion and considering the action of $\hat{U}^+_{6,2}(1)$ on $\hat{U}^-_{6,3}(2)$ using the above expression; and also dropping any partial derivative with respect to 1 and to 2 in the final expression since the lower point amplitude  $A_4(3^-,4^-,5^+,6^+)$ in \eqref{non_perturbative_factorization} doesn't depend on either of the two momenta, we obtain

\begin{equation}
\begin{aligned}
\hat{U}^+_{6,2}(1)\hat{U}^-_{6,3}(2)=& 1+\frac{\braket{21}}{\braket{26}} \bar{\lambda}^{\dot{a}}_1\frac{\partial}{\partial \bar{\lambda}^{\dot{a}}_6}  
+ \frac{[6(1+2)6\rangle}{[36]\braket{62}} {\lambda}^{{a}}_2\frac{\partial}{\partial {\lambda}^{{a}}_3} \\
&+\frac{\langle 6(1+2)3]}{\braket{26}[36]} {\lambda}^{{a}}_2\frac{\partial}{\partial {\lambda}^{{a}}_6}+\mathcal{O}(\epsilon^2)
\end{aligned}
\end{equation}

Combining these soft expansions together, we obtain

\begin{equation}
\begin{aligned}
\hat{S}_{6,2}(1^+)\hat{S}_{6,3}(2^-)=& \frac{1}{\langle 6(1+2)3] \left (2 p_6 \cdot q_{12} \right )}\left [
\frac{\braket{62}^2}{\braket{61}}\left (
[31]-\frac{\braket{62}[36][21]}{2 p_6 \cdot q_{12}}
\right)\right .\\
& +\frac{\braket{62}^3[36]}{\braket{61}\braket{12}}
+\frac{\braket{62}^3[36]}{ \braket{61}\braket{12}}
\left (\frac{\braket{21}}{\braket{26}} \bar{\lambda}^{\dot{a}}_1\frac{\partial}{\partial \bar{\lambda}^{\dot{a}}_6} 
+ \frac{2 p_6\cdot q_{12}}{[36]\braket{62}} {\lambda}^{{a}}_2\frac{\partial}{\partial {\lambda}^{{a}}_3} \right.\\
&\left .\left .+\frac{\langle 6(1+2)3]}{\braket{26}[36]} {\lambda}^{{a}}_2\frac{\partial}{\partial {\lambda}^{{a}}_6}\right )\right ]
+\mathcal{O}(\epsilon^0).
\end{aligned}
\label{first_expansion}
\end{equation}
We can follow the same steps to explore the soft expansion of the second term of the double soft operator \eqref{double_operator}, in which the non perturbative form is  

\begin{equation}
\begin{aligned}
\hat{S}_{1,3}(2^-)\hat{S}_{6,3}(1^+)
=&\frac{[13]^2}{[12][23]}\frac{[1(2+3)6\rangle}{\langle 6(1+2)3]}\frac{1}{s_{123}}\hat{U}^-_{1,3}(2)\hat{U}^+_{6,3}(1).
\end{aligned}
\end{equation}
The soft expansion around 1 and 2 of the above expression up to the next leading order leads to terms of order up to $1/\epsilon$ 
 
\begin{equation}
\begin{aligned}
\hat{S}_{1,3}(2^-)\hat{S}_{6,3}(1^+)=&\frac{1}{\langle 6(1+2)3] \left (2 p_3 \cdot q_{12}\right )}\left[
\frac{[13]^2}{[23]}
\left(\braket{26}-\frac{[13]\braket{36}\braket{21}}{2 p_3 \cdot q_{12}}
\right)\right.\\
&+\frac{[13]^3\braket{36}}{[12][23]}+\frac{[13]^3\braket{36}}{[12][23]}\left (\frac{[12]}{[13]} {\lambda}^{{a}}_2\frac{\partial}{\partial {\lambda}^{{a}}_3} 
+ \frac{2p_3\cdot q_{12}}{[13]\braket{36}} \bar{\lambda}^{\dot{a}}_1\frac{\partial}{\partial \bar{\lambda}^{\dot{a}}_6} \right.\\
&\left.\left.+\frac{\langle 6(1+2)3]}{\braket{63}[13]} \bar{\lambda}^{\dot{a}}_1\frac{\partial}{\partial \bar{\lambda}^{\dot{a}}_3}\right )\right ]+\mathcal{O}(\epsilon^0).
\end{aligned}
\label{second_expansion}
\end{equation}
Adding the contribution of the soft expansion \eqref{first_expansion} and \eqref{second_expansion}, we can extract the leading term of the double soft operator by collecting terms of order $1/\epsilon^2$. This order is consistent with two independent single soft operators. 
Then, the next leading order can be obtained by collecting terms of order $1/\epsilon$ and can keep going to collect higher order in the expansion. In that regard, the leading of the double soft operator is given by

\begin{equation}
\begin{aligned}
{S}_{6,3}^{(0)}(1^+,2^-)=& \frac{-1}{\langle 6(1+2)3]}\left (
\frac{1}{2 p_6 \cdot q_{12}}\frac{[63]\braket{62}^3}{\braket{61}\braket{12}}
-\frac{1}{2 p_3 \cdot q_{12}}\frac{\braket{63}[31]^3}{[12][23]}
\right ),
\end{aligned}
\end{equation}
and the next leading order is given by

\begin{equation}
\begin{aligned}
{S}_{6,3}^{(1)}(1^+,2^-)=&
\frac{\braket{62}^3[36]}{\langle 6(1+2)3] \braket{61}\braket{12}\left (2 p_6 \cdot q_{12}\right )}
\left (\frac{2p_6\cdot q_{12}}{[36]\braket{62}} {\lambda}^{{a}}_2\frac{\partial}{\partial {\lambda}^{{a}}_3} 
\right .\\
&\left . 
+\frac{\braket{12}}{\braket{62}} \bar{\lambda}^{\dot{a}}_1\frac{\partial}{\partial \bar{\lambda}^{\dot{a}}_6} 
-\frac{\langle 6(1+2)3]}{[36]\braket{62}} {\lambda}^{{a}}_2\frac{\partial}{\partial {\lambda}^{{a}}_6}  \right )
\\
&+\frac{[13]^3\braket{63}}{\langle 6(1+2)3] [12][23] \left (2 p_3 \cdot q_{12}\right )}
\left (  \frac{2p_3\cdot q_{12}}{[13]\braket{63}} \bar{\lambda}^{\dot{a}}_1\frac{\partial}{\partial \bar{\lambda}^{\dot{a}}_6}\right .\\
&\left .
-\frac{\langle 6(1+2)3]}{\braket{63}[13]} \bar{\lambda}^{\dot{a}}_1\frac{\partial}{\partial \bar{\lambda}^{\dot{a}}_3}
+\frac{[12]}{[31]} {\lambda}^{{a}}_2\frac{\partial}{\partial {\lambda}^{{a}}_3}\right )
\\&-\left (\frac{\braket{62}^2[16]}{\braket{61}\left(2 p_6 \cdot q_{12}\right )^2}+\frac{[31]^2\braket{32}}{[23]\left(2 p_3 \cdot q_{12}\right )^2}\right ).
\end{aligned}
\end{equation}

The expressions of leading and next leading soft we derived here match perfectly with the results from perturbative calculation made in \cite{Klose:2015xoa,Volovich:2015yoa, Georgiou:2015jfa}. The expressions from these works are directly derived from BCFW calculation under soft limit of the two momenta. While here, we first derive the non-perturbative expression of the double soft operator from the soft decomposition of split-helicity amplitude, then softly expand the expression so we can consistently check the non-perturbative soft operator.

\subsection{Triple soft}
\begin{figure}
\centering
\includegraphics[scale=.5]{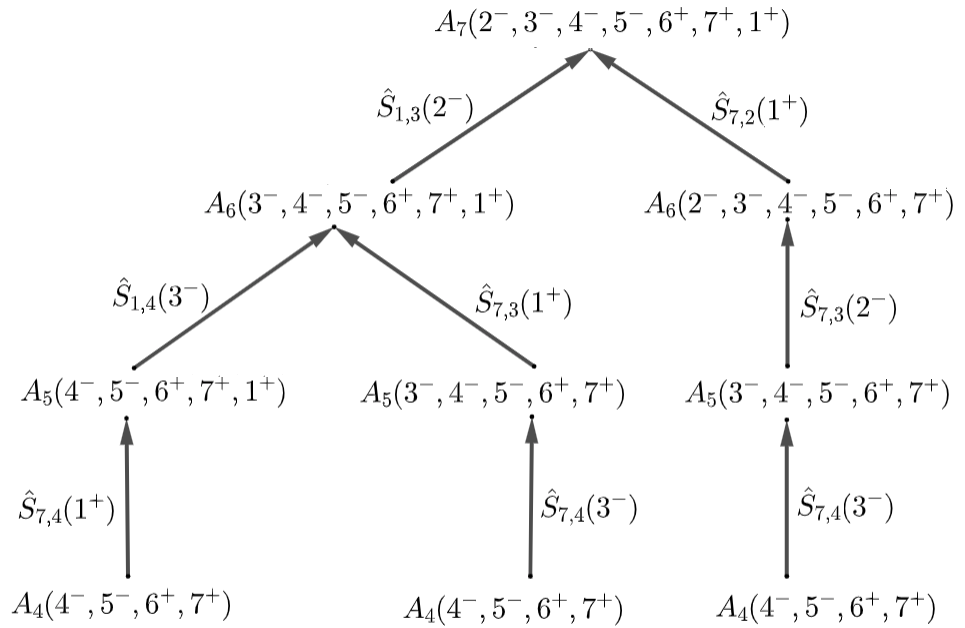}
\caption{Soft reconstruction of a seven point split-helicity amplitude.}
\label{A7_decomposition_tree}
\end{figure}

The analysis of the triple soft gluons in terms of soft reconstruction is entirely similar to that of double soft gluons described in the previous section. As before, we consider a split-helicity amplitude which will be successively decomposed using the soft reconstruction \eqref{inverse_soft_recursion_tree} into a reasonable term that reflects a factorization of the triple gluons contribution from a lower point amplitude. For such, we consider the seven point split-helicity 
\begin{equation}
A^\text{split}_{7,4}=A_7(2^-,3^-,4^-,5^-,6^+,7^+,1^+),
\end{equation}
where its soft diagram decomposition is given by fig.\ref{A7_decomposition_tree}. Such decomposition reflect the fact that the $A^\text{split}_{7,4}$ split-helicity can be reconstructed from a $A^\text{split}_{6,3}$ and a $A^\text{split}_{6,4}$ split-helicities which can be decomposed into lower points as in the derivation of the double soft. Analytically, such decomposition leads to the non-pertubative factorization
\begin{equation}
\begin{aligned}
 A_7(2^-,3^-,4^-,5^-,6^+&,7^+,1^+)=\\&\Big [\hat{S}_{1,3}(2^-)\hat{S}_{1,4}(3^-)\hat{S}_{7,4}(1^+)+\hat{S}_{1,3}(2^-)\hat{S}_{7,3}(1^+)\hat{S}_{7,4}(3^-)
 \\&
 +\hat{S}_{7,2}(1^+)\hat{S}_{7,3}(2^-)\hat{S}_{7,4}(3^-)\Big]
A_4(4^-,5^-,6^+,7^+),
\end{aligned}
 \end{equation}
where the $\hat{S}_{i,j}(k^\pm)$ are the single soft operator for gluon of $\pm1$ helicity given in \eqref{soft_operator}. From this factorization, we can identify the non-perturbative operator associated to triple soft gluon in terms of the individual single soft operator,
 \begin{equation}
 \begin{aligned}
  \hat{S}_{7,4}(1^+,2^-,3^-)=&\hat{S}_{1,3}(2^-)\hat{S}_{1,4}(3^-)\hat{S}_{7,4}(1^+)+\hat{S}_{1,3}(2^-)\hat{S}_{7,3}(1^+)\hat{S}_{7,4}(3^-)
 \\&
 +\hat{S}_{7,2}(1^+)\hat{S}_{7,3}(2^-)\hat{S}_{7,4}(3^-).
 \end{aligned}
 \label{triple_sof_operator}
 \end{equation}

As for the case of double soft gluons, we expand the triple soft operator as the three gluons ($1^+,2^-,3^-$) are taken to be soft. From the expansion, we can extract the leading contribution $S^{(0)}$ and the higher leading term $S^{(k)}$ from power counting in $\epsilon$. 

In terms of soft shift operators given in \eqref{soft-shifts}, we have
\begin{equation}
\begin{aligned}
 &\hat{S}_{1,3}(2^-)\hat{S}_{1,4}(3^-)\hat{S}_{7,4}(1^+)= \frac{[14]^2\langle7(2+3+4)1]}{[13][32][24]\langle7(1+2+3)4]}\frac{\hat{U}^-_{3,4}(2)\hat{U}^-_{1,4}(3)\hat{U}^+_{7,4}(1)}{s_{1234}}  ,
 \end{aligned}
 \label{triple_soft_d1}
\end{equation}

\begin{equation}
\begin{aligned}
 &\hat{S}_{1,3}(2^-)\hat{S}_{7,3}(1^+)\hat{S}_{7,4}(3^-)=\\
&\hspace{3cm}\frac{\langle7(2+3)1]^2[1(2+3)(1+2+3+7)4]}{[12][23][31]\langle7(1+2)3]\langle7(1+2+3)4]}\frac{\hat{U}^-_{1,3}(2)\hat{U}^+_{7,3}(1)\hat{U}^-_{7,4}(3)}{s_{123}\,s_{1237}},
 \end{aligned}
 \label{triple_soft_d2}
\end{equation}
and 
\begin{equation}
\begin{aligned}
 &\hat{S}_{7,2}(1^+)\hat{S}_{7,3}(2^-)\hat{S}_{7,4}(3^-)=\frac{\braket{27}^3[74]}{\braket{71}\braket{12}[34]\langle7(1+2)3]}\frac{\hat{U}^+_{7,2}(1)\hat{U}^-_{7,3}(2)\hat{U}^-_{7,4}(3)}{s_{127}},
 \end{aligned}
 \label{triple_soft_d3}
\end{equation}
where $s_{1237}=(p_1+p_2+p_3+p_7)^2$ is the generalized Mandelstam variable with four momenta, and $[i(j)(k)l]=[ij]\langle j(k)l]$ a spinor contraction as define in \cite{Britto:2004ap}. Following the same steps as in the double soft, we can derive the expression of the leading triple soft by expanding terms in \eqref{triple_soft_d1}, \eqref{triple_soft_d2} and \eqref{triple_soft_d3} as the three momenta ($1^+,2^-,3^-$) are taken to be soft. After collecting terms of order $1/\epsilon^3$ in the soft expansion and with some simplification, we find

\begin{equation}
\begin{aligned}
{S}_{74}^{(0)}(1^+,2^-,3^-)&=\frac{1}{\langle7(1+2+3)4]}\left(\frac{1}{2p_7\cdot q_{123}}\frac{\langle7(2+3)1]^3[74]}{[12][23]\langle7(1+2)3]  s_{123}}\right .
\\ &+\left.\frac{1}{2p_4\cdot q_{123}}\frac{[14]^3\braket{74}}{[13][32][24]}\right )+\frac{1}{2p_7\cdot q_{12}}\frac{\braket{27}^3[74]}{\braket{71}\braket{12}[34]\langle7(1+2)3]},\\[.3cm]
\end{aligned}
\label{leading_triple}
\end{equation}
with $q_{123}= p_1+p_2+p_3$.

The expression of leading soft we derived here is very compact and matches perfectly with the BCFW results calculated perturbatively in \cite{Volovich:2015yoa}. It can be checked that in successive soft, where 1 is softer than 2 which is softer than 3, the above result reduces into the products of triple single soft factors. While in the case where 3 is softer and 1 and 2 goes soft simultaneously, the expression is nicely reduce to a product of a single soft factor and a double soft factor. 

\section{Conclusion}
The aim of this work is to investigate the possibility of describing multiple soft operators in terms of single soft using soft decomposition. To achieve that goal, we start by showing how a split-helicity amplitude can be constructed from only two lower split-helicity amplitudes, which represents general way to softly decompose split-helicity. From such decomposition, we then demonstrate how double and three soft gluons can non-perturbatively be factorized respectively for helicity configuration $(1^+2^-)$ and $(1^+2^-3^-)$. We find that the associated non-perturbative soft operators, \eqref{double_operator} and \eqref{triple_sof_operator}, are fully expressed in terms of the individual single soft. Finally to test our results, we use soft expansion of the operators to recover the leading and sub-leading soft factor derived from standard BCFW calculation.

We show that the soft decomposition of split-helicity amplitudes is a powerful tool for generating multiple soft operators beyond perturbative calculations. For practical purposes of these operators, it is useful to generalize the soft decomposition for any amplitude using the non-perturbative multiple soft operator.

\end{document}